**$H_3^+$ and UKIRT – a saga of discovery ***

Tom Geballe

Gemini Observatory, Hilo, HI 96720

March 26, 2009

In 1981, about the time I was moving from the Carnegie Institution in Pasadena to UKIRT in Hilo, I received a phone call from Takeshi Oka, a professor in the Departments of Chemistry and Astronomy at the University of Chicago, a man I had not met and whose work I did not know. He asked if I would help him search dense interstellar clouds for the molecular ion $H_3^+$, a substance I also had never heard of and the astrophysical significance of which I obviously was completely unaware.

Oka had a deep knowledge of chemistry and physics, but little experience in observational infrared astronomy. He had gotten my name from my graduate advisor, Prof. Charles H. Townes, from whom I had received my PhD in physics at Berkeley seven years earlier. I was not all that experienced either, but I had helped to build several infrared spectrographs for astronomy. One of them had just been used by Caltech graduate student Daniel Nadeau, his advisor, Prof. Gerry Neugebauer, and me to make some significant discoveries regarding the recently detected shock-excited molecular hydrogen ($H_2$) in the Orion Molecular Cloud.

Oka's enthusiasm was such that despite my ignorance of $H_3^+$ there was no way I could turn him down. He had succeeded where others had failed in measuring the infrared spectrum of $H_3^+$ in the laboratory and was now eager to detect it in space. Indeed he had already attempted to do so at Kitt Peak National Observatory, but his attempt had been, in his words, a "miserable failure."

After a little investigation, the significance of $H_3^+$ in astronomy began to become apparent to me. $H_3^+$, an equilateral triangular arrangement of three protons, with two orbiting electrons, is the catalyst for nearly all of interstellar gas phase chemistry. Formed by the cosmic ray induced ionization of $H_2$ (the dominant species in molecular clouds) followed by the ion-molecule reaction, $H_2^+ + H_2 \rightarrow H_3^+ + H$, the newly created $H_3^+$ will gladly donates its spare proton to nearly any other neutral atom or molecule it encounters. The receiver of the proton then itself becomes highly reactive. Neutral-neutral chemistry is exceedingly slow in the low temperature environment of dark clouds, but ion-molecule reactions that include such proton hops are rapid and the reaction rates are nearly insensitive to temperature. Hence one can envision interstellar gas phase chemistry as a network of ion-molecule reactions that starts with the creation of $H_3^+$. Most of the molecular species that we observe in dense clouds, including CO, $H_2O$, $H_2CO$, $NH_3$, $CH_4$, and a number of cyanopolyynes such as $HC_5N$, $HC_7N$, and $HC_7N$ that Oka and his collaborators had recently discovered using radio telescopes, were thought to be created via this network.

Ion-molecule chemistry as the dominant chemistry in molecular clouds had been proposed in the 1970s, and the idea had been quickly accepted. But $H_3^+$, its smoking gun, had not been detected. Presumably it was rare. Yes, it was being created at a good clip; a typical hydrogen molecule in a typical dark cloud – 0.1 parsec in extent and containing maybe $3x10^{56}$ $H_2$ molecules - had to wait "only" about a billion years, $3x10^{16}$ seconds, to be ionized by the cosmic ray background. Because it is the rate at which this ionization occurs that controls the production of $H_3^+$, the rate of total production in such a cloud is simply the ratio of those two numbers. Thus, about$10^{40}$ $H_3^+$ molecules would be created in the cloud each second. This seems like an astoundingly large rate and it seems that the process should produce an even more astoundingly large total number of $H_3^+$ molecules in the cloud, and it does. But $H_3^+$ is highly reactive; it destroys itself any time it bumps into just about anything other than hydrogen or helium. So even though that total number is huge, the cloud is vast and the steady state concentration of $H_3^+$ in it is predicted to be relatively low, typically a few parts per billion. This meant that the spectroscopic signature of interstellar $H_3^+$ likely would be very faint.

As I pondered how to best search for $H_3^+$, whose strongest transitions are in its fundamental asymmetric stretching (vibrational) band in the 3-4μm region, a strategy soon emerged: perform absorption spectroscopy of a dense molecular cloud using as a light source, an infrared-bright young stellar object in or perhaps behind the cloud. The two requirements (in addition to a good infrared telescope on a good infrared site) were a sensitive high-resolution spectrograph capable of measuring weak and narrow absorption lines at those wavelengths and a sufficiently bright stellar object to provide high signal-to-noise ratio. Together these might make it possible to detect the expected weak lines of $H_3^+$. Unfortunately both were in short supply. UKIRT was not due to receive such an instrument for a few years, no other sensitive 3-4μm spectrograph existed on Mauna Kea, and based on the expected sensitivity at UKIRT of the new instrument (a combination cold grating and warm Fabry-Perot spectrometer) one could count the sufficiently bright and suitable embedded stellar sources on the fingers of one's hands.

When the aforementioned spectrometer became available at UKIRT, Oka and I successfully applied for observing time. During the mid and late 1980s we made a few attempts to detect $H_3^+$, all of which were unsuccessful. Some of the upper limits to the abundance of $H_3^+$ that we obtained were close to the predicted values and in one cloud a faint and almost believable signature of $H_3^+$ was present. But it was clear that, assuming ours was the best observing strategy, significant improvement would require the next generation of infrared spectrometers.

At about the time Oka's and my paper reporting these negative results appeared in the literature, news came of the exciting and totally unexpected discovery, made at the Canada-France-Hawaii Telescope, that the $H_3^+$ overtone band at 2μm produces bright high altitude aurorae at the poles of Jupiter. For the moment the two of us set aside our interest in the interstellar medium in order to follow up at UKIRT on this first detection of $H_3^+$ outside the laboratory. On several mornings during September of 1989, after the night's observers departed, we successfully sought and measured many fundamental band lines of $H_3^+$ in Jupiter.

These observations, which we published in 1990, are a rare example of daytime science done at UKIRT. I recall that each morning as Oka and I took over the telescope, the seeing was excellent. However, a little over an hour after we had started to observe the entire dome and telescope, and the observers, would begin to shake. The construction crew building the paved road connecting the summit road and the NASA Infrared Telescope and Keck telescopes had arrived and had started its work; the crew was excavating directly down the hill from UKIRT! We persevered through that until a little after 8:00 a.m., when the seeing would suddenly degrade to a few arc seconds, presumably due to onset of ground layer turbulence in the air surrounding the telescope, caused by the heating of the loose lava on the summit. Our view of Jupiter became very blurry. That change brought an end to each morning's observing.

We were excited to finally be seeing $H_3^+$ in space. Our observations showed that the lines of the fundamental band of $H_3^+$ were impressively strong in Jupiter's hydrogen-rich atmosphere and that, although they were brightest very close to the poles, they were also present elsewhere. Within a few years we and our collaborators, in particular Steve Miller of University College London and Laurence Trafton of the University of Texas, now using the newly arrived and revolutionary two-dimensional (58x62) array spectrometer, CGS4, were able to show that the Jovian $H_3^+$ line emission extended across the entire planet. We also discovered that on Uranus $H_3^+$ line emission also is present across the entire planet and that Saturn possesses prominent $H_3^+$ aurorae. We did not succeed in detecting $H_3^+$ on Neptune, and nobody else has succeeded to date.

By then it was the mid-1990's and Oka's and my attention turned back to the interstellar medium and the hunt for $H_3^+$ in dense clouds. The next generation of sensitive high-resolution infrared spectrographs on Mauna Kea was here, in the form of CGS4 at UKIRT and CSHELL at the NASA Infrared Telescope Facility (IRTF). At the time the spectral resolution achievable with CSHELL was 2-3 times higher than CGS4. Although we were obviously quite familiar with UKIRT and CGS4 (and I was by then Head of Operations at UKIRT), we were concerned that CGS4's spectral resolution was not nearly high enough to be optimal for detection of the expected weak and narrow lines of interstellar $H_3^+$. Therefore in autumn 1994 we applied for observing time at the IRTF. Our proposal was rejected. The same thing happened the following semester. Along with the rejections came some suggestions from the time allocation committee of how we might improve our proposal. The next semester we submitted what we believed was an improved proposal. To our dismay we were turned down a third time.

Meanwhile at UKIRT, CGS4 was undergoing an upgrade to its current 256x256 array. The new array had much smaller pixels, giving CGS4 the capability of matching CSHELL in spectral resolution. Although we had missed the most recent semester deadline for UKIRT proposals, I knew that there was a mechanism for quickly obtaining a small amount of observing time, the UKIRT Service Observing program. It had also become clear to me that due to the cold, high, and usually dry atmosphere above Mauna Kea, we could probably make a better choice of lines to initially seek; a close pair of $H_3^+$ lines near 3.67μm should give us a somewhat better chance for detection than the single slightly stronger line at 3.95μm that we had used in our earlier searches. Soon our proposal for a couple of hours of Service time to search for $H_3^+$ in two dense clouds was accepted and on April 29, 1996 I found myself at UKIRT as the scheduled (by me!) observer on a Service night.

Late that night, within a few minutes of beginning the observation of the first cloud, I could see the evidence for $H_3^+$ in the form of two weak but apparently real absorption lines. Within another hour the second cloud showed a somewhat more convincing detection. I excitedly emailed Oka the news. We decided to be cautious. Two and one-half months later, on July 15, we repeated the observations, using as a convenient test the changing Doppler shift of the lines caused by the orbital motion of the earth around the sun. The lines were there again, more clearly than in April and slightly shifted in wavelength, just as expected (see Fig. 1). Analysis showed that the amount of $H_3^+$ we had detected was consistent with that predicted from the ion molecule chemical model. The smoking gun had been detected and the viability of the ion-molecule model was confirmed. By the end of November we had published a Letter in the journal, Nature, announcing this. As can be seen in Fig. 1, we experienced our fifteen minutes of fame. It had been fifteen years since we started on our quest.

This is only a part of the story of $H_3^+$ at UKIRT, the telescope at the center of my career as an astronomer, and the Joint Astronomy Centre, an observatory whose uniquely supportive staff, past and present, remains dear to my heart. UKIRT has played the seminal role in both the discovery and exploration of $H_3^+$ in the interstellar medium, as well as contributing enormously, via initial detections and detailed investigations, to our understanding of $H_3^+$ in the gas giant planets of the outer solar system. As of the end of 2008 Oka and I together with our collaborators had published more than twenty refereed papers on $H_3^+$ based entirely or in part on observations made at UKIRT. (There are still a few more papers in our heads, waiting to be written.) In addition to the initial detection of interstellar $H_3^+$ and the initial detections on Uranus and Saturn, the unexpected discovery that $H_3^+$ is abundant in the diffuse interstellar medium, a finding that has had far reaching implications, occurred at UKIRT. It also should be noted that the first detection of $H_3^+$ in the interstellar medium of an external galaxy was made at UKIRT.

Many key questions involving the astrophysics of $H_3^+$ require telescopes larger than UKIRT for answers. However, there remains plenty of important $H_3^+$ science for 4-meter class telescopes, and particularly those on Mauna Kea. Should UKIRT's excellent Cassegrain spectrographs return to action Oka and I will happily and eagerly request to use them to continue our research.

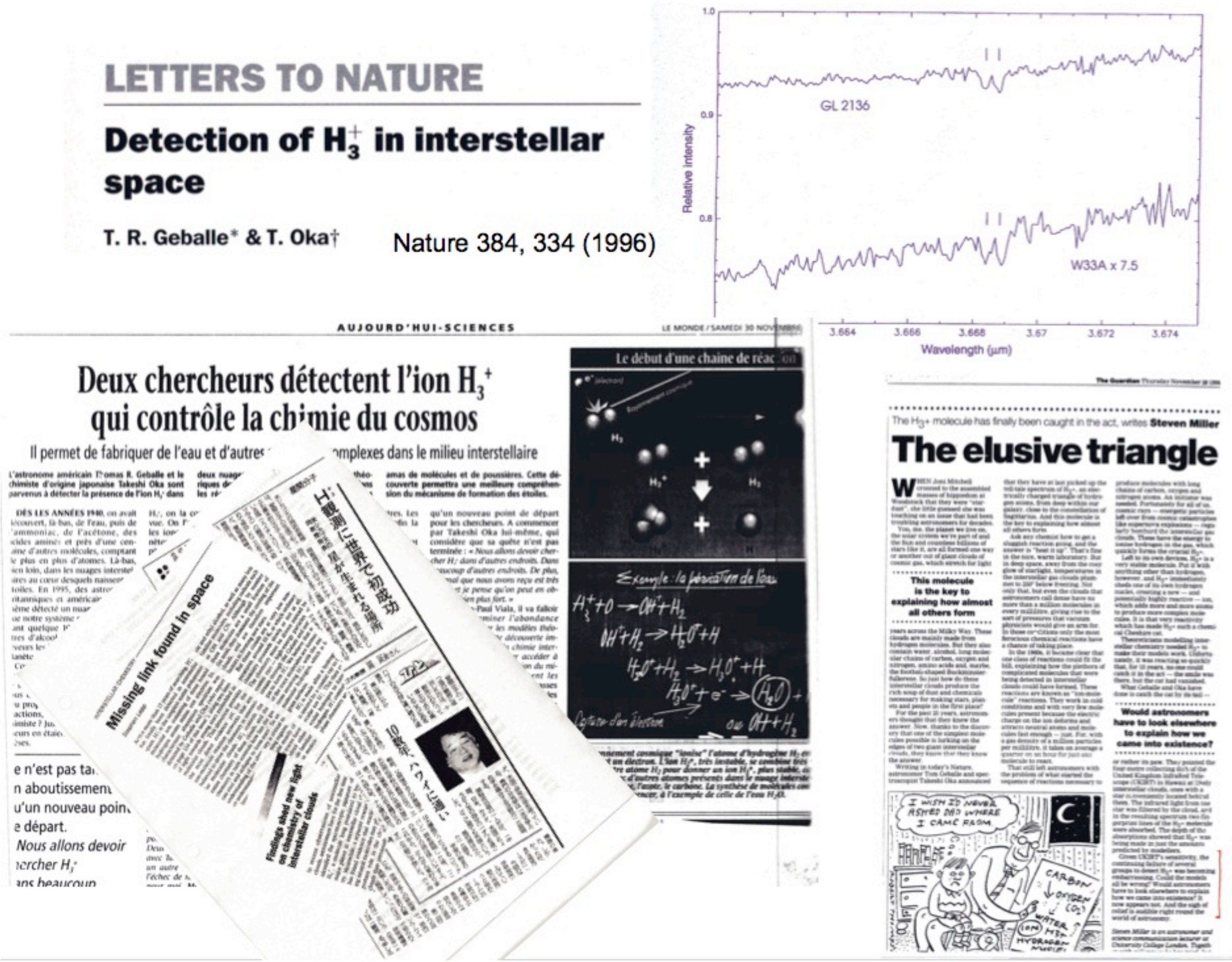

Figure caption – At top left is part of the title page of the discovery paper published in Nature on 28 November 1996. At top right are the spectra of the two molecular clouds from July 15, 1996 that clinched the detection of $H_3^+$. The predicted wavelengths of the absorption lines in each cloud are indicated by the short vertical lines. At bottom is some of the reaction of the world press.